\documentclass[twocolumn,trackchanges]{aastex61}

\usepackage{comment}
\usepackage{amsmath}
\submitjournal{AJ}

\shorttitle{Origin of Neptune's unusual satellites}
\shortauthors{Li \& Christou}

\begin{document}
\title{The origin of Neptune's unusual satellites from a planetary encounter}

\correspondingauthor{Daohai Li}
\email{lidaohai@gmail.com, li.daohai@astro.lu.se}

\author[0000-0002-8683-1758]{Daohai Li}
\affiliation{Lund Observatory\\Department of Astronomy and Theoretical Physics\\Lund University, Box 43, SE-221 00 Lund, Sweden}

\author{Apostolos A. Christou}
\affiliation{Armagh Observatory and Planetarium\\ College Hill, Armagh, BT61 9DG \\ Northern Ireland, UK}

\begin{abstract}

The Neptunian satellite system is unusual, comprising Triton, a large ($\sim2700$ km) moon on a close-in, circular, yet retrograde orbit, flanked by Nereid, the largest irregular satellite ($\sim$300 km) on a highly eccentric orbit. Capture origins have been previously suggested for both moons. Here we explore an alternative in-situ formation model where the two satellites accreted in the circum-Neptunian disk and are imparted irregular and eccentric orbits by a deep planetary encounter with an ice giant (IG), like that predicted in the Nice scenario of early solar system development. We use $N$-body simulations of an IG approaching Neptune to 20 Neptunian radi ($R_\mathrm{Nep}$), through a belt of circular prograde regular satellites at 10-30 $R_\mathrm{Nep}$. We find that half of these primordial satellites remain bound to Neptune and that 0.4-3\% are scattered directly onto wide and eccentric orbits resembling that of Nereid. With better matches to the observed orbit, our model has a success rate comparable to or higher than capture of large Nereid-sized irregular satellites from heliocentric orbit. At the same time, the IG encounter injects a large primordial moon onto a retrograde orbit with specific angular momentum similar to Triton's in 0.3-3\% of our runs. While less efficient than capture scenarios \citep{Agnor2006}, our model does indicate that an in-situ origin for Triton is dynamically possible. We also simulate the post-encounter collisional and tidal orbital evolution of Triton analogue satellites and find they are decoupled from Nereid on timescales of $\sim$$10^4$ years, in agreement with \citet{Cuk2005}.

\end{abstract}

\keywords{Satellite formation --- 
Neptunian satellites --- Irregular satellites --- Close encounters -- Celestial mechanics}

\section{Introduction}\label{sec-intro}

\begin{figure}
\centering
\includegraphics[width=\hsize]{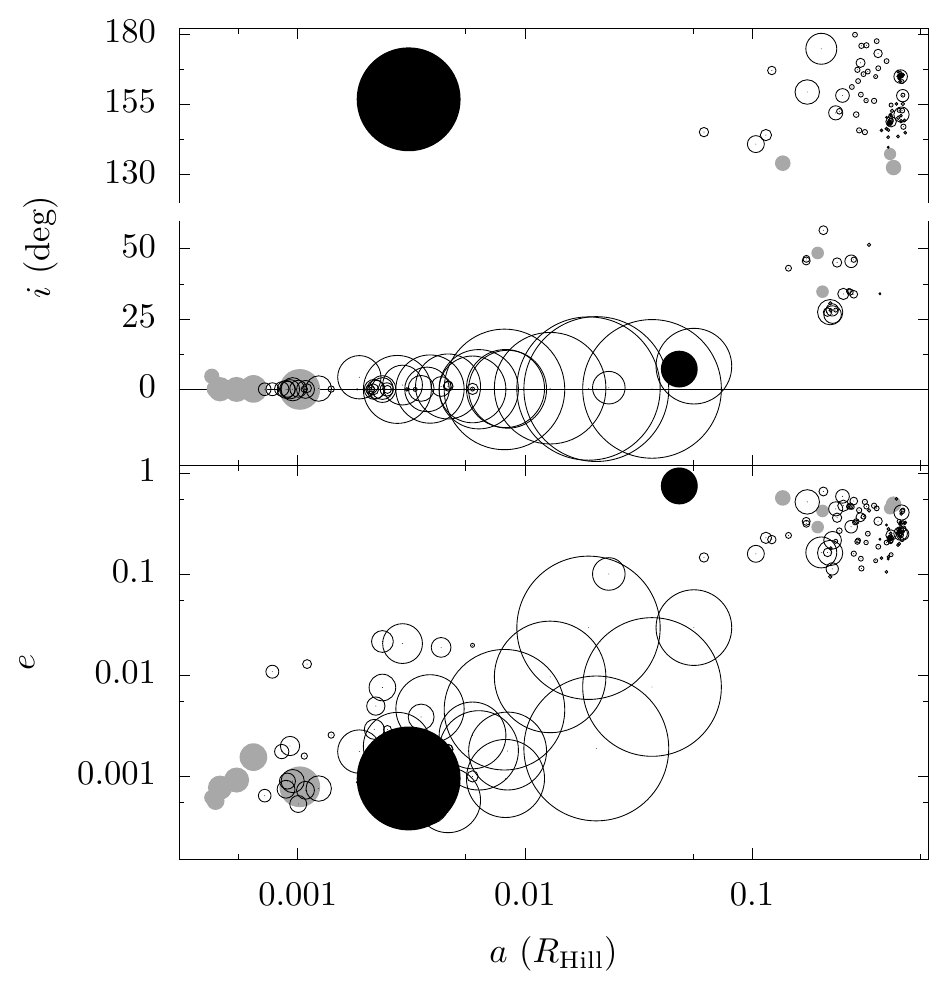}
\caption{Distribution of giant planet moons in the $(a,i)$ and $(a,e)$ planes from Scott Sheppard's website:  \url{https://sites.google.com/carnegiescience.edu/sheppard/moons}. Symbol size is proportional to the square root of that moon's actual size. Grey symbols represent moons around Neptune while Triton and Nereid are the large and small black filled circles respectively. Small (5-10$\times 10^{-4}$) vertical offsets have been added to allow moons with negligible orbital eccentricity to be displayed on a log-scale.}
\label{fig-aei-size-new-hill}
\end{figure}

Observational biases notwithstanding, Neptune has the least number of satellites among the four giant planets but perhaps with the most intriguing orbits. The largest moon, Triton, is orbiting its host planet at 14 Neptunian radii ($R_\mathrm{Nep}$) on a circular path but, oddly, in a retrograde direction. Nereid, $>200R_\mathrm{Nep}$ further out and the third largest moon in the system, has the highest orbital eccentricity among solar system moons (Figure \ref{fig-aei-size-new-hill}).

Mechanisms favouring capture of Triton from heliocentric orbit include gas drag \citep{McKinnon1984,McKinnon1995}, collisions \citep{Goldreich1989} and three-body gravitational interaction \citep{Agnor2006,Nogueira2011,Vokrouhlicky2008}. See \citet{Colombo1971,Heppenheimer1977,Pollack1979,Cuk2004d,Nicholson_etal_2008,Nesvorny2007,Nesvorny2014} for discussions on satellite capture.

In an alternative {\it in situ} formation scenario, the two moons have accreted in the circum-Neptunian disk \citep[e.g.,][]{Szulagyi2018} with initially circular, prograde orbits on the equatorial plane of Neptune. This raises the question of how they arrive at their {\it current} orbits.

\citet{Harrington1979} originally postulated an encounter between an {\it ad hoc} planetary body of several earth masses and Neptune, flipping Triton's orbit and scattering Nereid outward. This scenario has been criticized \citep{Farinella1980,McKinnon1995a} on the grounds that the encountering planet is not observed in the solar system and that the encounter may have over-excited Neptune's orbit. Also, computational resources available at that time allowed only one ``successful'' simulation run, making it difficult to estimate the success rate of this particular evolutionary path. In a subsequent model where Triton was assumed captured, \cite{Goldreich1989} suggested that Nereid could be scattered onto a wide orbit by Triton, an outcome not reproduced in numerical simulations \citep{Nogueira2011}.

It is believed the giant planets radially migrated in the early solar system \citep{Fernandez1984,Malhotra1993} in the now widely-accepted framework of the Nice model. There, the planets formed at different heliocentric distances from those where they are presently observed and, due to interactions with a primordial planetesimal disk, they migrated to their current locations. Since its introduction \citep{Tsiganis2005}, the Nice model has evolved considerably to meet an enhanced set of constraints. Because of the difficulty to correctly excite the orbit of Jupiter \citep{Morbidelli2009} and, at the same time, to avoid over-exciting the inner main asteroid belt \citep{Morbidelli2010a,Minton2011} and the terrestrial planets \citep{Brasser2009,Agnor2012}, Jupiter is thought to have impulsively ``jumped'' over the 2:1 mean motion resonance with Saturn, owing to close encounters with an ice giant. As such, a five-planet variant of the Nice model, where an additional ice giant planet (IG hereafter) was subsequently ejected from the solar system, was introduced \citep{Batygin2012a,Nesvorny2012}. The IG, before its ejection, probably encountered other planets as well, leading to the capture of Trojans and irregular satellites \citep{Nesvorny2013,Nesvorny2014} and the emplacement of the so-called ``kernel'' of the Kuiper Belt \citep{Nesvorny2015}. These planet-planet encounters may have been as close as 0.003 au \citep{Nesvorny2014a}, penetrating to the satellite region.

The appearance of the Nice scenario mitigates the two major objections \citep{Farinella1980,McKinnon1995a} to the {\it in-situ} formation of Triton \citep{Harrington1979} since the encountering IG could have been ejected from the solar system and thus be rendered unobservable while Neptune's eccentricity, even if excited, may be damped owing to the interaction with the planetesimal disk (actually, the encounters considered here ensure that the orbit of Neptune is at most mildly excited). Hence it is worthwhile to reexamine the {\it in-situ} formation model within the constraints of the Nice scenario, also providing statistics of successful vs unsuccessful simulations runs to estimate model efficiency and exploring the ensuing Neptunian system evolution post-encounter. We focus on studying how a close encounter between Neptune and the IG could bring about the unusual orbits of Triton and Nereid. Our model consists of three parts. First (Section \ref{sec-enc}), we check how the IG encounter scatters an initial population of Neptunian moons onto distant, eccentric orbits as well as onto retrograde orbits. In Section \ref{sec-col} we study the system's post-encounter evolution via collisions and tidal dissipation. The implications and conclusions are presented in Sections \ref{sec-dis} and \ref{sec-con}.

\begin{figure}
\centering
\includegraphics[width=\hsize]{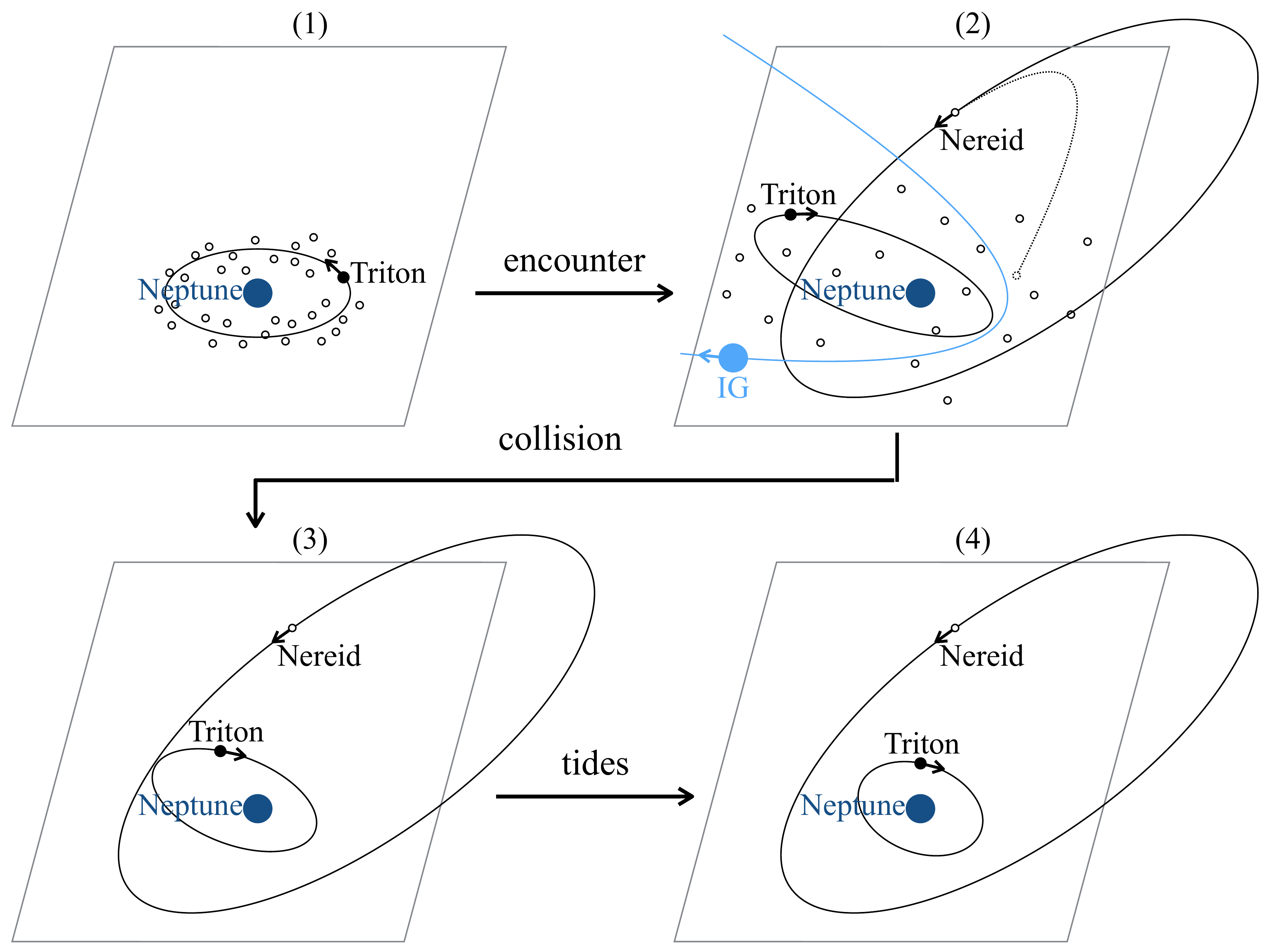}
\caption{Illustration of our origin scenario for Triton and Nereid. Before the encounter, several tens of pre-existing moons orbit Neptune (1), all small (open circles) except one that is Triton-sized (black filled circle). After the IG encounter (2), the orbit of the large moon is flipped and a small moon is emplaced onto a wide, eccentric orbit, turning into Nereid. Subsequently (3), collisions with Triton remove the other small moons and decouple the orbit of Triton from Nereid's. Finally (4), tides circularise Triton's orbit.}
\label{fig-illustration}
\end{figure}

\section{NEPTUNE-IG encounter}\label{sec-enc}
Following \citet{Cloutier2015} and as detailed below, we first set up the ice giant (IG)-Neptune encounters; then the satellites' evolution under these encounters are examined. See panels (1) and (2) of Figure \ref{fig-illustration} for an illustration of this phase.

\subsection{Model}
We consider a three-body system comprising the Sun, the IG and Neptune. In order to fully control the minimum separation $d_\mathrm{enc}$ between the IG and Neptune, we start with the Sun-IG-Neptune system at the moment of the two planets' closest approach \citep{Cloutier2015}, i.e., when the relative position vector satisfies $|\vec r_\mathrm{IG-N}|=d_\mathrm{enc}$; we set $d_\mathrm{enc}=$0.003 au \citep[or 18 Neptunian radii][]{Nesvorny2014a}. At this moment, the relative velocity vector is perpendicular to the relative position vector: $\vec v_\mathrm{IG-N} \bot \vec  r_\mathrm{IG-N}$. We further assume their relative kinetic energy $|\vec  v_\mathrm{IG-N}|^2$ to be uniformly distributed within the range $(0,3  v^2_\mathrm{esc}$) where $ v_\mathrm{esc}$ is the two-body escape velocity between the IG and Neptune. The orientations of the two vectors are random in the solid angle. Then the IG-Neptune barycentre is assigned a heliocentric orbit parameterised by the set of orbital elements  $(a_\mathrm{IG+N},e_\mathrm{IG+N},i_\mathrm{IG+N})$, with values within the ranges: $a_\mathrm{IG+N}\in(10,40)$ au, $e_\mathrm{IG+N}\le0.7$ and $i_\mathrm{IG+N}\le10^\circ$; the angular elements are randomly drawn from a uniform distribution. We calculate the position and velocity vectors $\vec r_\mathrm{IG+N}$ and $\vec v_\mathrm{IG+N}$ and combine them with $\vec r_\mathrm{IG-N}$ and $\vec v_\mathrm{IG-N}$ to fully define the heliocentric state vectors of the IG and Neptune at the instant of closest approach. Next, we carry out a reference frame transformation such that the $z$-axis is parallel to the total angular momentum of the three-body system. We refer to this as the heliocentric frame. The IG mass is 18 Earth masses \citep{Nesvorny2014a}, similar to that of Neptune.

We integrate this system backwards and forwards for 10 years apiece using the general Bulirsch-Stoer algorithm in {\small MERCURY} \citep{Chambers1999} with an error tolerance of $10^{-14}$. At the end of these two integrations, we check the planets' mutual distance is larger than 0.8 au, Neptune's current Hill radius (see below). Then the heliocentric orbital elements of Neptune are used to determine whether an encounter is ``realistic'' in that its semimajor axis is in the range 25 au$<a_\mathrm{N}<$30 au and eccentricity is $e_\mathrm{N}<0.15$; For the inclination, we only require that the change in the direction of the heliocentric angular momentum vector before and after the encounter is $\Delta i_\mathrm{N}<6^\circ$, for consistency with Nice scenario simulations \citep{Nesvorny2012,Gomes2018} and with constraints derived from Neptune's perturbation on the Cold Kuiper Belt Objects (CKBOs) \citep{Wolff2012,Dawson2012}. We additionally require that the change in the Neptunian semimajor axis before and after the encounter is $<$1 au \citep{Nesvorny2015}. While a more violent dynamical history of Neptune is not necessarily inconsistent with the CKBOs \citep{Morbidelli2014,Gomes2018}, the planets' exact heliocentric orbits are irrelevant for the satellites' evolution during the brief encounter. Furthermore, this encounter need not be the last one and subsequent encounters may additionally change the orbit of Neptune. Thus the Neptunian orbit in our simulations may not define the starting point of ensuing outward migration and damping in $e$ and $i$, upon which the constraints from CKBOs were imposed \citep{Dawson2012}.

\begin{figure}
\begin{center}
\includegraphics[width=\hsize]{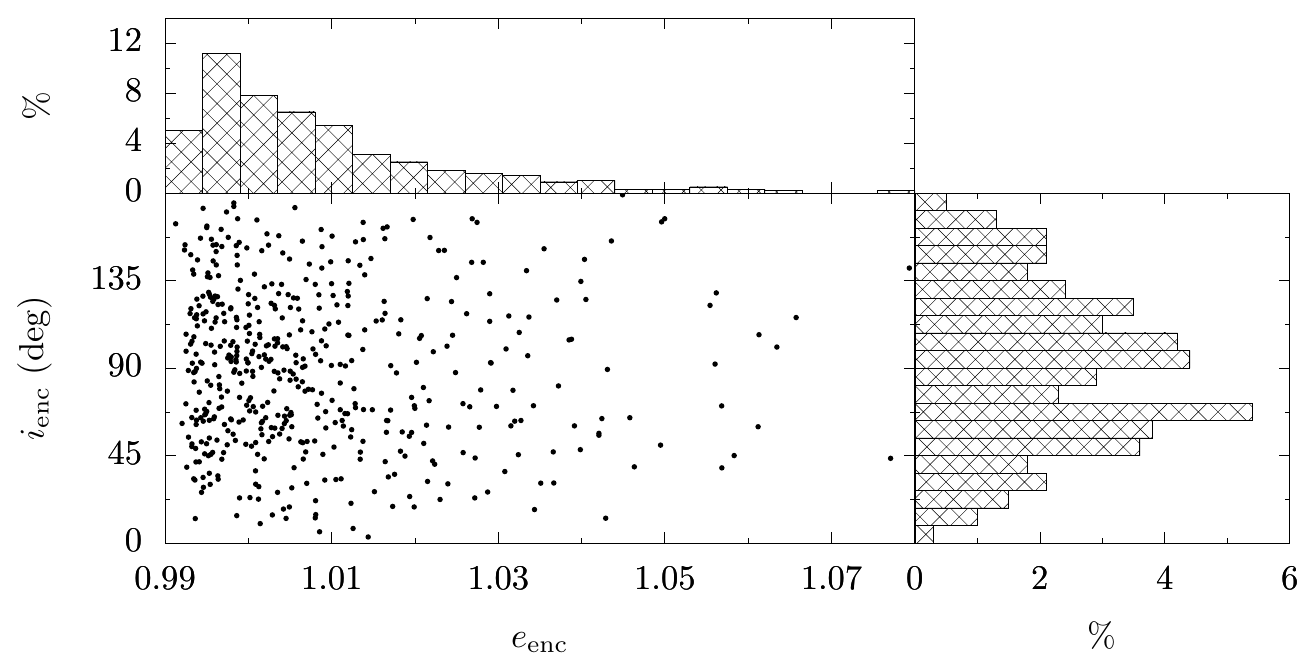}
\caption{Distribution of eccentricity and inclination of the IG in the Neptune-centric frame, the latter measured wrt Neptune's equator. Histograms of the two quantities are shown in the top and right panels.}
\label{fig-IG-nep-ei}
\end{center}
\end{figure}

A total of 500 encounters are generated in this way. In Figure~\ref{fig-IG-nep-ei}, we show the distribution of the IG's orbit with respect to Neptune at closest approach. As expected \citep[cf.][]{Deienno2014}, they are mostly near-parabolic due to strong gravitational focusing. The resemblance of the distribution of $i_\mathrm{enc}$ (measured in the Neptunian-centric frame, see below) to a sine function suggests that the encounters are nearly uniformly distributed in the solid angle. Hence, the direction of Neptune's spin axis is statistically unimportant.

Then we generate 1000 test moons on prograde circular orbits, all coplanar with respect to the Neptunian equatorial plane (the Neptune-centric frame), with orbits evenly distributed in the range $(10 R_\mathrm{Nep},30 R_\mathrm{Nep})$. The orbital phase is again randomly drawn. Since the planetary spin axis is essentially unaffected by close encounters \citep{Lee2007}, we assume that Neptune acquires its current obliquity of $\sim 30^\circ$ before the encounter \citep[by, for example, a giant impact,][]{Morbidelli2012b}.

The Sun, IG, Neptune and 1000 test moons are integrated for 20 yr using {\small MERCURY} where the moons are treated as massless particles. During the integration, a moon is removed if it collides with either the IG or Neptune. After the integration, we calculate the orbital elements of the moons in the Neptune-centric frame. A test moon is removed from the simulation if its semimajor axis exceeds half the Neptunian Hill radius \citep{Nesvorny2003} $R_{\rm Hill}=a_\mathrm{N}{\left(M_\mathrm{N}/M_{\rm Sun}\right)}^{1/3}$ ($M_{\rm Sun}$ is the solar mass) or if it achieves a hyperbolic orbit. Moons remaining on bound orbits around Neptune are referred to in the following sections as survivors.

\subsection{Results}
\begin{figure}
\centering
\includegraphics[width=\hsize]{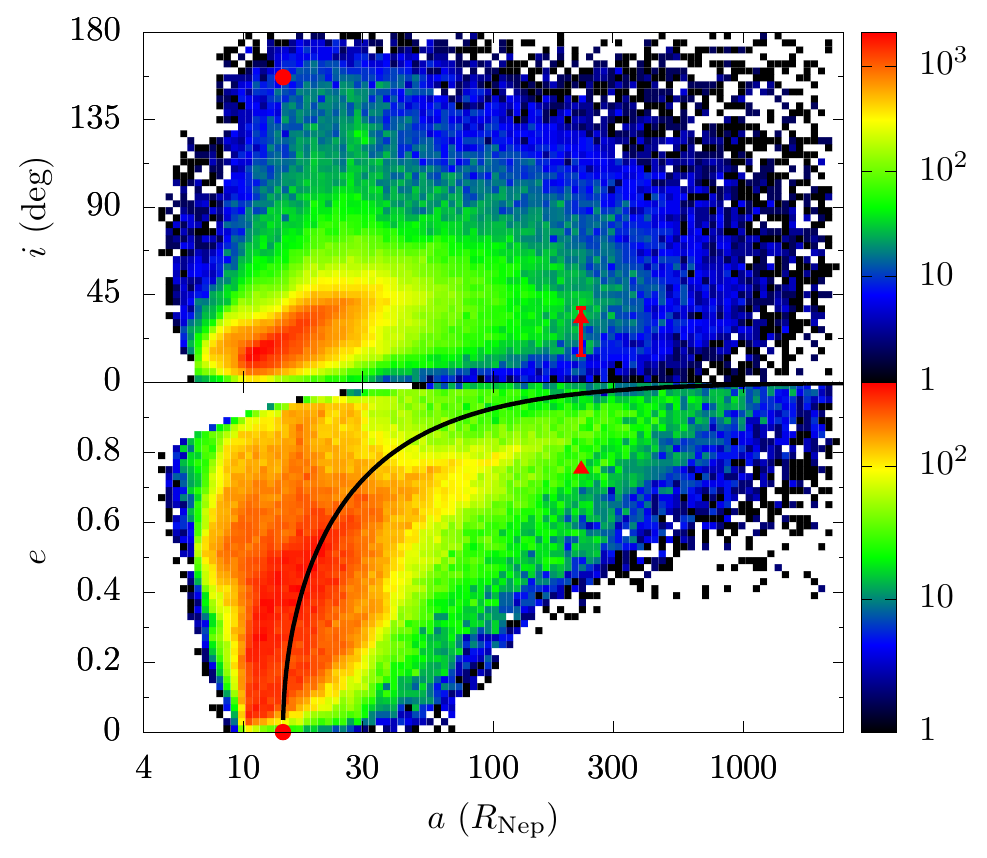}
\caption{Density histogram of all $2.7\times10^5$ surviving test particles in Phase 1 that remain bound to Neptune after the IG flyby in the $(a,i)$ and $(a,e)$ planes. Warmer colours represent higher values. The red points \& triangles respectively mark Triton's \& Nereid's present orbit and the error bar shows Nereid's inclination variational range. Note that it is those orbits on the right of Triton's (thus with larger $a$) that finally evolve towards it later, following equal-angular momentum level curves (black curve in bottom panel).}
\label{fig-a-e-i}
\end{figure}

Out of our 500$\times$1000=500000 test moons, 53\% survive and their orbital distribution is shown in Figure \ref{fig-a-e-i}. Not unexpectedly, most have acquired significant eccentricities and inclinations. A small fraction gain orbits with semimajor axes greater than about $100R_\mathrm{Nep}$, eccentricities up to unity and inclinations up to $180^\circ$. Specifically, we observe that the orbits of both Triton (red circle) and Nereid (red triangle) lie within the distribution of simulated moons.

\begin{figure*}
\begin{center}
\includegraphics[width=0.8\hsize]{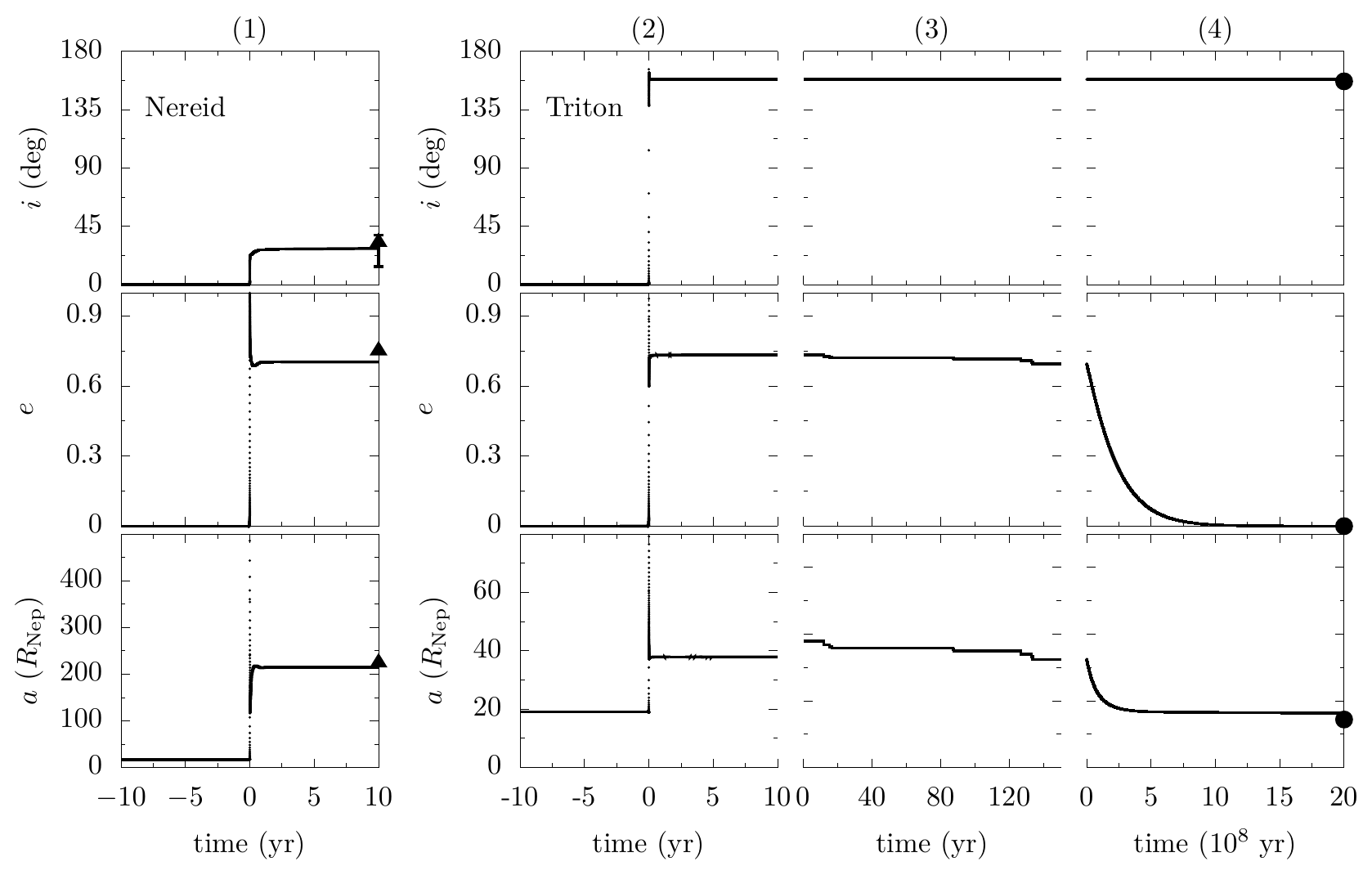}
\caption{Time evolution of the orbital elements for a Nereid Analogue (NerA) and a Triton Analogue (TriA). At time zero (columns 1 and 2) the IG is closest to Neptune. During the encounter, the NerA is directly transferred to a Nereid-like wide and eccentric orbit (1) while, at the same time, the orbit of the TriA is flipped and its eccentricity excited (column 2), causing it to intersect Nereid's. Subsequently, collisions with the other surviving moons rapidly decrease the $a$ and $e$ of the TriA (column 3) while preserving the NerA. Finally, tides circularise the TriA's orbit in $\sim$ Gyr (column 4). Large triangles and circles represent the actual Nereid and Triton.}
\label{fig-aei-tot}
\end{center}
\end{figure*}

To determine how well the orbits of the two moons can be reproduced in Phase 1 simulations, we need to quantify how closely a test moon orbit from the simulation should resemble the orbit of the actual satellite. For Nereid ($a=224 R_\mathrm{Nep}$, $e=0.75$ and $i=32^\circ$) we consider those particles injected onto orbits with $a\in (200R_\mathrm{Nep},250R_\mathrm{Nep})$ as Nereid Analogues (NerAs). The time evolution of a typical NerA is shown in column (1) of Figure~\ref{fig-aei-tot}. During the encounter, the NerA is instantly scattered onto a wide, highly eccentric and inclined orbit, analogous to that of the observed satellite (black triangles on the right).

\begin{figure*}
\begin{center}
\includegraphics[width=0.8\hsize]{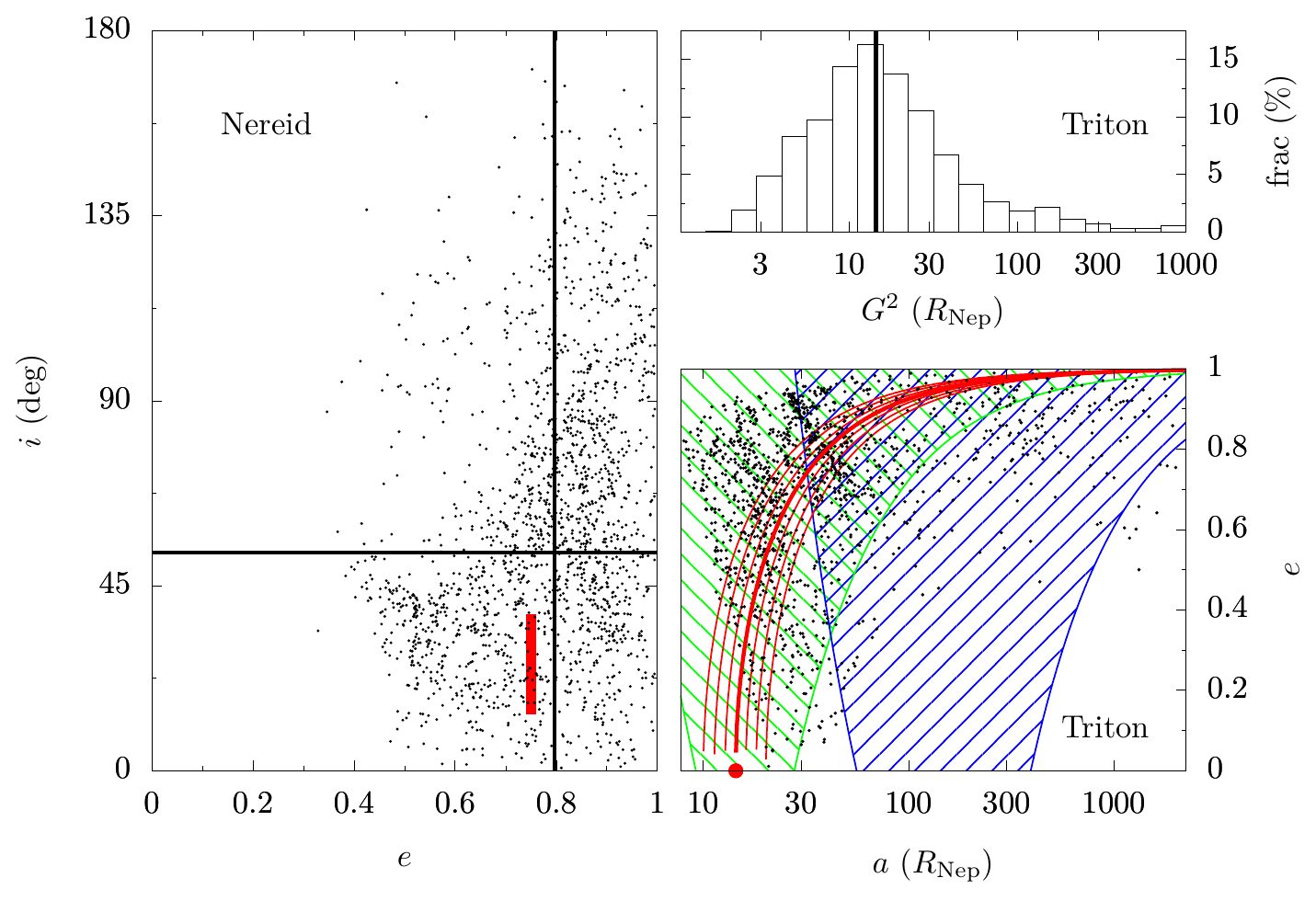}
\caption{Orbital distribution of the Nereid Analogues (NerAs) and Triton Analogues (TriAs) obtained from the encounter simulations. Left: distribution of all the $1.8\times10^3$ NerAs in the $(e,i)$ plane. The variations of Nereid's actual $i$ and $e$ are shown as the red regions. Black vertical and horizontal lines represent the median $e$ and $i$ of all NerAs. Bottom right: distribution of all the $1.5\times10^3$ TriAs in the $(a,e)$ plane. Triton is the red circle. A TriA in the blue region may collide with Nereid while one in the green region may collide with the other surviving small moons (see text). The red curves are level curves of angular momentum, with the thick one corresponding to that of Triton (red point). During collisional and tidal evolution, angular momentum and inclination are quasi-constant and TriAs evolve along the red curves. Top right: histogram of the square of the normalised angular momentum $G^2=a(1-e^2)$ of all TriAs. The median is 14.1 $R_\mathrm{Nep}$, $\sim$2\% away Triton's value 14.4 $R_\mathrm{Nep}$ (black vertical line).}
\end{center}
\label{fig-tri_ner}
\end{figure*}

From the simulations we obtain $1.8\times10^3$ such NerAs or about $1.8\times10^3/5\times10^5\sim0.4\%$ of the initial prograde satellite population. Because NerAs are defined through $a$, we show their $e$ and $i$ distribution in the left panel of Figure \ref{fig-tri_ner}. Here the red region marks the range of eccentricity variation for Nereid and Triton while the vertical and horizontal lines are the median values obtained from NerAs. While the agreement in eccentricity is excellent, the obtained median inclination of the NerAs is somewhat higher but still brackets the observed value to within $1-\sigma_i$ ($\simeq 34^\circ$).

Next, we turn our attention to Triton. For reference, Triton has $a=14.4 R_\mathrm{Nep}$, $e\approx0$ and $i=157^\circ$. We define test moons on orbits with $i\in(150^\circ,165^\circ)$ as Triton Analogues (TriAs). A typical time evolution of a TriA is presented in column (2) of Figure~\ref{fig-aei-tot}.

We obtain $1.5\times10^3$ TriAs or about $1.5\times10^3/5\times10^5\sim0.3\%$ of the initial population. Their distribution in $a$ and $i$ is shown in the bottom right panel of Figure \ref{fig-tri_ner}. Unlike the NerAs matching the observed orbit fairly well, the TriAs immediately after the IG encounter, in general, have wider and more eccentric orbits than the real Triton. Thus, as with Triton's post-capture evolution, additional mechanisms must be invoked to both shrink and circularise the orbit \citep{Goldreich1989,McKinnon1995,Cuk2005,Correia2009,Rufu2017}. We explore these in the next Section.

\section{Aftermath: Evolution of circumneptunian material post-encounter}\label{sec-col}
Following Neptune's encounter with the IG, the surviving satellites will undergo further evolution in the form of mutual gravitational interactions, collisions between each other and with Neptune as well as tidal decay of the orbits. The outcome of this phase must satisfy the observational constraints i.e. the survival of one Triton-sized and one Nereid-sized moon. Gravitational perturbations leading to planetary impact and, to a lesser extent, collisional elimination remove small satellites in Triton-crossing orbits over $10^{4} - 10^{5}$ yr \citep{Cuk2005,Nogueira2011}, requiring an efficient protection mechanism for Nereid. 

We begin to tackle this by considering the ``soft'' constraint that, for all solar system giant planets, the ratio of the satellites' total mass to that of the host planet, is $\lesssim 0.024\%$ \citep{Canup2006,Barr2017}. As Triton itself is already $\sim0.02\%$ of the mass of Neptune, it is reasonable to expect exactly one large, Triton-sized moon with the remaining mass of surviving moons (Other Surviving Small Moons or OSSMs) being small enough so that Triton survives the ensuing collisional evolution. It has been argued that a head-on impact with an impactor mass of no more than a few \% of Triton's mass would not disrupt Triton \citep{Rufu2017}. The total mass of OSSMs must satisfy the constraint that $\Sigma m_\mathrm{OSSM} \lesssim m_\mathrm{Triton}$ which may be converted into a rough estimate of their number. For instance, if each of the OSSMs is assumed to be of Nereid's mass ($\sim$0.14\% of that of Triton; \url{https://ssd.jpl.nasa.gov/?sat_phys_par}), there should be no more than several hundred. This is discussed further in Section~\ref{sec-dis}.

We now consider the effect of OSSM-Triton collisions on Triton's orbit. We want to find out how Triton's orbit is altered and at what rate. For this purpose, a system comprised of Neptune, the TriA and a user-defined number of OSSMs is integrated with {\small MERCURY}. We run these simulations with 200 OSSMs per TriA, at the high end of the estimated non-Triton mass orbiting Neptune. However, the sole purpose of this exercise is to find out the critical mass  - hopefully much lower than the threshold for disruption - of OSSMs needed for orbital decoupling such that the apocentre distance of the TriA becomes smaller than the pericentre distance of the NerA. And, as we will see, the model is not dependent on this particular size frequency distribution of the OSSMs so long as their mass is at least a few percent of Triton's.

A TriA is assumed to be of Triton's size and mass \citep{Murray1999} and its starting orbit is taken from the encounter simulations. Each OSSM, assumed to be of Nereid's size and mass, is assigned the median of all prograde orbits from the encounter simulations, a random orbital phase and a random orbit orientation. Mutual gravitational interactions as well as the solar perturbation are omitted \citep{Nogueira2011}, leaving the Neptunian quadrupole term parameterised by the $J_2$ coefficient \citep{Murray1999}. Collisions are treated as perfect mergers with the change in the TriA's orbit calculated via conservation of linear momentum. The integration is terminated as soon as the TriA is decoupled from Nereid's orbit or the simulation reaches $10^4$ yr, irrespective of whether decoupling occurs or not.

Not all TriAs need to be examined. As the bottom right panel of Figure~\ref{fig-tri_ner} shows, only TriAs in the green region can collide with the OSSMs. An orbit outside the green region has a pericentre distance larger than the apocentre distance of OSSMs while an orbit outside of the blue region does not pose a risk to the NerA since the orbits do not intersect. Hence, only the $4.9\times10^2$ TriAs within the intersection of the blue and green regions need to be considered.

An example run is shown in column (3) of Figure~\ref{fig-aei-tot}. In this case, orbit decoupling is achieved over 150 yr after 8 collisions. The timescale is irrelevant; for our purposes, the essential feature of the evolution is that the collisions reduce both $a$ and $e$ and  leave $i$ unchanged. From our numerical runs, we find that the TriA's orbit is collisionally decoupled from the NerA's in 96\% of cases, typically after after colliding with a mass of about 2.7\% of Triton's mass (or with about 19 of the OSSMs in the model). 

\begin{figure}
\begin{center}
\includegraphics[width=\hsize]{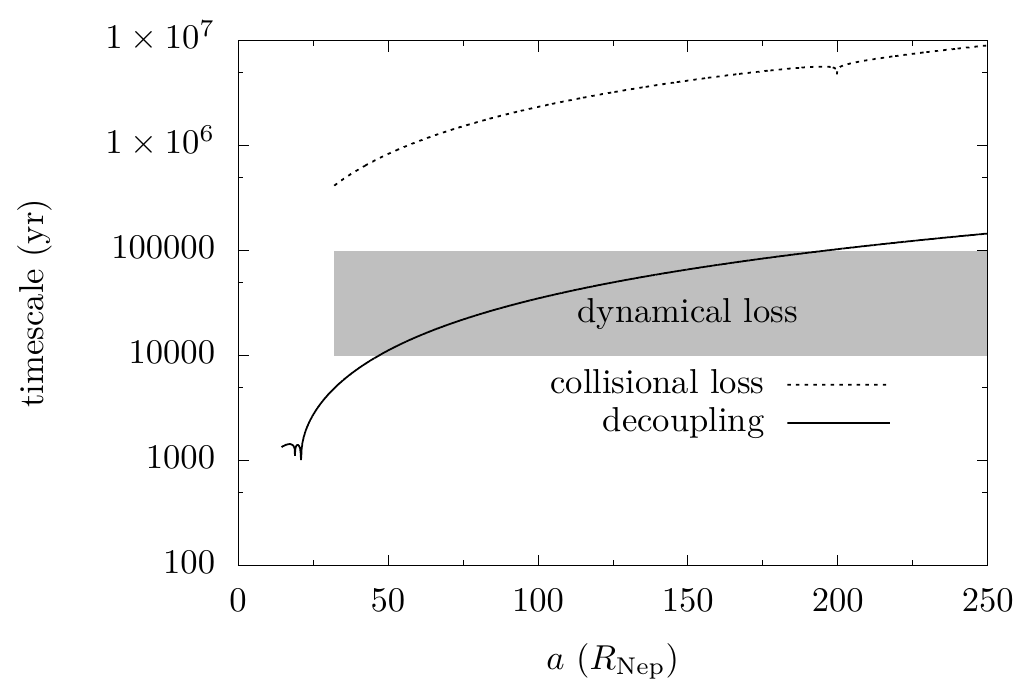}
\caption{Collisional (dotted line) and dynamical loss (shaded region) timescale for Nereid compared to the decoupling timescale for Triton (solid line). For $a$ sufficiently small, the orbit of the TriA no longer crosses that of the NerA.}
\label{fig-collision}
\end{center}
\end{figure}

We calculate analytically the collisional timescale $T_\mathrm{TriA,OSSM}$ between an OSSM and a TriA using the approach of \citet{Kessler1981}. We estimate the time to collide with 2.7\% of Triton's mass in OSSMs using a truncated harmonic series and find that the time to decouple Triton's orbit is approximately $3.5 T_\mathrm{TriA,OSSM}$ when there are about 20 OSSMs; this is the decoupling timescale. The solid line in Figure~\ref{fig-collision} shows this timescale as a function of the TriA's semimajor axis. Note that this timescale is weakly dependent on the number of bodies that the mass is divided into (i.e., $\propto \log(N)$) and varies from about $3.5 T_\mathrm{TriA,OSSM}-8.2 T_\mathrm{TriA,OSSM}$ when the number of bodies is changed from 20 to 2000 while keeping the total mass constant. Be it $3.5 T_\mathrm{TriA,OSSM}$ or $8.2 T_\mathrm{TriA,OSSM}$, for semimajor axes less than about $100 R_{Nep}$, it is not more than a few times $10^{4}$ years \citep[see also][]{Cuk2005,Rufu2017}, comparable to the loss timescale for dynamical interactions by Triton \citep[$\sim10^4-10^5$ yr, shaded region,][]{Cuk2005,Nogueira2011}. Therefore, the conditions arising after the Neptune-IG encounter favour the survival of  Nereid against either dynamical or collisional elimination as long as the TriA acquires an orbit $\lesssim 100 R_\mathrm{Nep}$. This is, in fact, true for the vast majority of TriA orbits arising from the encounter phase (Figure \ref{fig-tri_ner}, bottom right panel).

Therefore, collisions bring down Triton's orbit efficiently enough to preserve Nereid \citep[as well as any other moons on wide orbits,][]{Cuk2005} and eliminate the OSSMs. As the typical total mass of OSSMs colliding with Triton is only a few \% of Triton's, this moon's angular momentum and inclination remain effectively unchanged.

Our assumption that all OSSMs have the same mass as Nereid is not essential for collisional damping of Triton's orbit. Rather, our model requires that there are some other surviving moons after the IG encounter, and these moons amount to at least a few $\sim\%$ the mass of Triton. Also, the OSSM orbits must be relatively close to Neptune (cf. Figure \ref{fig-a-e-i}) such that $T_\mathrm{TriA,OSSM}$ is smaller than the dynamical lifetime of Nereid.

Following the depletion of the OSSM population, tidal dissipation within the satellites and within the planet further shrinks and circularises the orbits (panels 3 and 4 of Figure \ref{fig-illustration}).

Nereid's orbit is too far from Neptune to be significantly affected by tides and is not considered further in the following.
Triton's tidal evolution has been discussed extensively in the literature \citep{McKinnon1984,Chyba1989,Goldreich1989,Cuk2005,Correia2009,Nogueira2011,McKinnon1995a}. Here we want to know specifically the effect of tides on the orbit of the TriA following collisional evolution. For this purpose, we follow a recent implementation \citep{Correia2009} of the equilibrium tidal model \citep{Hut1981}. Since the orbit of a TriA is now entirely inside that of Nereid, orbital precession is controlled by Neptune's oblateness \citep{Goldreich1989,Nogueira2011,Li2016} and the solar perturbation may be omitted. Actually, we can also  disregard the planetary oblateness, as it causes orbital precession but not secular variations in $a$, $e$ or $i$ \citep{Nogueira2011}. We consider the TriA as a rocky body with Love number $k_\mathrm{Tri}=0.1$ and tidal $Q_\mathrm{Tri}=100$ \citep{Goldreich1989,Correia2009,Nogueira2011}; the solid body parameters for  Neptune are taken from \citet{Correia2009,Hubbard1991}.

A typical TriA evolution is shown in column (4) of Figure~\ref{fig-aei-tot}. As with collisions, $i$ is unaffected while both $a$ and $e$ decrease and the orbital angular momentum is quasi-conserved. Here the orbit of the TriA is circularised within a Gyr, consistent with other studies \citep{Goldreich1989,Correia2009,Nogueira2011}. This timescale is shorter if the TriA is molten or semi-molten \citep{McKinnon1984,Goldreich1989,Nogueira2011}. In fact, for any TriA with pericentre distance $q<20 R_\mathrm{Nep}$, tides circularise the orbit within the age of the solar system \citep{Nogueira2011}.

\section{Discussion}\label{sec-dis}

\subsection{Comparison with observations and model efficiency}
We first examine how well our model matches the observed orbits of Triton and Nereid.

Since its orbit is circular, Triton is defined by its orbital $i$ and $a$. In our model, both $i$ and the normalised angular momentum $G$ are quasi-conserved during collisional as well as tidal evolution and are thus determined solely from the IG-encounter phase. By definition, all TriAs have inclinations close to Triton's, therefore we only need to consider $G$. In the top right panel of Figure \ref{fig-tri_ner}, we show that the median of $G^2$ of our TriAs is within 2\% of the observed value, suggesting that our model reproduces the correct final orbit for Triton.

On the other hand, a NerA is directly placed onto a wide, highly eccentric orbit during the encounter phase and experiences no further evolution. By definition, NerAs all have $a$ similar to Nereid's. As shown in the left panel of Figure \ref{fig-tri_ner}, the median $e$ of the NerAs is 0.80, $\sim6\%$ from that observed. The dispersion in $i$ of the NerAs is large and Nereid is within $1-\sigma_i$.

The efficiency of our model is determined mainly by the Neptune-IG encounter phase because the outcome of subsequent evolution -  producing a Triton-like object with orbit circularised within the age of the solar system - is fairly deterministic and the NerA does not participate in the latter phase.

The simulation results indicate that about $1.8\times10^3$ of the initial $5\times10^5$ prograde satellite particles ended up on Nereid-like orbits after the encounter with the ice giant and are labelled as NerAs.  This suggests that the probability of an object with Nereid's orbit and size resulting from a similarly deep encounter is $P_{NerA} \sim 4\times10^{-3} N_{R>R_\mathrm{Nereid}}$, where $N_{R>R_\mathrm{Nereid}}$ is the number of Nereid-sized moons ($R_\mathrm{Nereid}\simeq170$-km) in the initial satellite system. If there were a few tens of these moons $N_{R>R_\mathrm{Nereid}}\sim20$ in the primordial system, the probability of an encounter producing a NerA is about $P_\mathrm{NerA}\sim0.1$.  Similarly, when examining the encounter results we find that $1.5\times10^3$ of the $5\times10^5$ initial prograde satellite particles are transferred to orbits with Triton-like retrograde orbits and are labelled as TriAs.  This suggests that the probability of an object with Triton's inclination and size resulting from a deep encounter is about $P_\mathrm{TriA}\sim 3\times10^{-3}N_{R>1000\text{-km}}$, where $N_{R>1000\text{-km}}$ is the number of large ($R>1000$-km) satellites in the initial prograde satellite population. Another factor, the chance of a 0.003 au encounter happening, is $\sim0.1$ \citep{Deienno2014,Nesvorny2014a}. Hence, the overall success rate of our model to account for both Triton and Nereid in this way is about $3\times10^{-5}$.

Clearly, how stringently we define TriAs and NerAs has a great impact on the model efficiency. If we only ask the TriAs to have an inclination $i>90^\circ$ instead of $i\in(150^\circ,165^\circ)$, the chance for creating one such object increases by a factor of ten to $3\%$. Similarly, if all orbits with $a>100R_\mathrm{Nep}$ are recognised as NerAs rather than requiring $a\in(200R_\mathrm{Nep},250R_\mathrm{Nep})$, the corresponding rate also rise by an order of magnitude, reaching also $3\%$. Combined, this suggests that if the analogues are loosely defined, the overall efficiency reaches a few times $10^{-3}$.

\subsection{Comparison with capture models}
In-situ formation models for the two moons have been discussed in Section~\ref{sec-intro}, so here we focus on capture models.
A leading mechanism for Triton's capture is via three-body encounter
\citep{Agnor2006}. In this model, Triton and a bound massive binary companion encounter Neptune.  During this encounter the orbit of the binary is tidally disrupted and leaves the Triton-mass object on a bound orbit around Neptune, while its companion escapes on a hyperbolic trajectory.
The capture efficiency, examined in the context of the Nice scenario, has been estimated to be between 2\% \citep{Vokrouhlicky2008} and 50\% \citep{Nogueira2011}; though, this exchange capture may have occurred before the Nice scenario with a higher efficiency \citep{Vokrouhlicky2008}. Hence, in terms of Triton's procurement alone, our model is less likely than capture via 3-body gravitational encounters \citep[][]{Agnor2006,Vokrouhlicky2008,Nogueira2011}.

However, Nereid's acquirement is also non-trivial. Nereid is the largest among the so-called irregular satellites (Figure \ref{fig-aei-size-new-hill}) and larger than Trojan asteroids, populations genetically linked in that they were both captured by the giant planets during the instability period from the primordial planetesimal disk \citep[PPD,][]{Nesvorny2013,Nesvorny2007,Nesvorny2014}. The largest Jovian Trojan (624) Hektor and irregular moon (J VI) Himalia have been used to show the consistency between capture efficiencies and the size frequency distribution (SFD) and the total mass of the PPD \citep{Nesvorny2016}. However, Nereid does not readily fit into this picture. For example, Hektor, the second largest object within the two populations, is 230 km in diameter \citep{Nesvorny2016} and its capture efficiency, as a Trojan at Jupiter, is $(6-8) \times10^{-7}$ \citep{Nesvorny2013}, $\sim 20$ times higher than the capture of irregular satellites at Neptune \citep{Nesvorny2014}. Furthermore, the steep SFD \citep{Nesvorny2016} implies Nereid-sized (or larger) objects are rarer than Hektor by almost an order of magnitude. These facts combined suggest that the capture of Nereid is $\sim 100$ times as infrequent as that of Hektor: should capturing one Hektor-sized object on Jovian Trojan orbits be expected \citep{Nesvorny2016}, acquiring one Nereid-sized body at Neptune must be unlikely. Indeed, following these works, the expected number of such large objects captured at Neptune is $< 0.006$.

In summary, the chance of capturing both Triton \citep{Agnor2006} and Nereid \citep{Nesvorny2007} is thus $10\% \times 0.6\%\sim6\times10^{-4}$, higher than the average efficiency of our model by a factor of ten. We note, capture models do not strongly constrain the characteristics of the capture orbit. Leaving Triton aside, Nereid has the largest orbital eccentricity and the smallest semimajor axis (in units of the host planet's Hill radius) among all irregular moons (Figure \ref{fig-aei-size-new-hill}). Indeed, Nereid is located at the inner edge of the region where the capture mechanism operates \citep{Nesvorny2007,Nesvorny2014}. As discussed before, our efficiency reaches a few times $10^{-3}$ if we loosen the requirement on the orbital similarity between the observation and the analogues and this is higher than that of the capture models by a factor of a few.

However, the origin of the two moons may not be related at all. For example, Triton may be captured long before that of Nereid and when the latter arrives, the orbit of the former has already been small enough -- no issue for the stability of Nereid. Then it is perhaps unfair to compare the overall efficiency between the in-situ and the capture models and only the individual rate should be confronted. As discussed before, the exchange capture model for Triton is probably better while our model seems to work particularly well for Nereid. So can the two work together? Timing is important. (1) If Triton predates Nereid: Upon capture, Triton gains a highly eccentric orbit of which the circularisation would probably have cleared any small moons, leaving no seeds for Nereid anymore. Also, Triton itself may be excited or even ejected by the encounter. (2) Or if Nereid precedes Triton: During the encounter, Nereid was placed onto its current orbit with some other small moons surviving. Then when Triton is captured, it collides with these other moons and Nereid is protected. So it seems that (2) may be viable but a careful modelling is needed.

\subsection{The primordial satellite population}
The scenario advocated here, operates with an efficiency of $\sim10^{-5}$ in a self-contained and self-consistent way, reproducing the main orbital features of both Triton and Nereid. Here we discuss its principal weaknesses.

However, how realistic is the assumed size distribution of the initial satellite population? This has been constructed based on the known inventory of solar system moons and our model efficiency (Section \ref{sec-enc}).

Currently, Jupiter and Uranus each has four large, similar-sized moons plus smaller ones. \footnote{But we do note that such four-moon systems may turn into ones containing a single large moon plus others \citep{Asphaug2013}.} Perhaps the system most closely resembling our assumed configuration is that of Saturn where the mass budget is dominated by Titan together with several intermediate-sized moons (each of $0.5\%-1.7\%$ of Titan's mass; \url{https://sites.google.com/carnegiescience.edu/sheppard/moons/saturnmoons}) but only a few, hundred km-sized moons. So, small number statistics aside, it seems that our assumption on the initial size distrbution of Neptune moons does not have concrete observational support. However, available to us are only the satellites' {\it current} configurations: e.g., moons exterior to Uranus' outermost moon Oberon could have been lost \citep{Deienno2011} and may not necessarily be primordial.

While the assumption that one Triton-sized moon exists seems sensible at least mass-wise \citep{Barr2017,Szulagyi2018}, the number of smaller moons, a few tens, is estimated from the expectation that one Nereid should be created during the best encounters (those featuring the highest occurrence rates of TriAs \& NerAs). A few tens of such moons, under these encounters, give rise to a NerA at an efficiency close to unity and, fortuitously, provide just the right amount of impacting mass to shrink the orbit of the TriA (its creation still a small-likelihood event) quickly enough to protect the NerA without disrupting the TriA. Nonetheless, when estimating the number of small moons using the overall creation efficiency for the NerA ($\sim$0.1\%), an initial population of a few hundreds results. These  moons, if each of Nereid's mass, total $\sim10\%$ of that of Triton. This population, while more efficiently decoupling Triton from Nereid, its large total mass implies a considerable decrease in the normalised angular momentum $G$ of Triton. Then, our argument about the agreement between the observed value and simulations fails (top right panel of Figure \ref{fig-tri_ner}). On the other hand, if only a few moons exist before the encounter it would be difficult to create a NerA and its survival becomes problematic.

Finally, we comment on the implications for Neptune's remaining moons. The irregular satellites are omitted from our discussion as these moons may have been acquired by Neptune later on \citep{Nesvorny2007}. The inner regular moons, however, will be perturbed by the IG. To quantify the maximum possible extent of the perturbation, we consider Proteus, the inner neighbour of Triton at 4.7 $R_\mathrm{Nep}$. For each of our 500 encounters, we place 10 test moons at 5 $R_\mathrm{Nep}$ around Neptune and follow them through the encounter. We observe that all these $500\times10=5000$ moons are stable and their orbital excitation is small, with a median eccentricity is $\lesssim0.04$. Moreover, because this moon is prograde and outside the synchronous orbit, it must have migrated outwards in the past. Therefore, Proteus has been closer to Neptune during our encounter and should have been disturbed to an even smaller degree. This experiment represents a worst-case-scenario for the disturbance to the inner regular satellites. Yet, these moons may also be perturbed by Triton, gaining moderate eccentricities which accelerates the rate of internal tidal dissipation \citep{Rodriguez2011} or leads to disruptive collisions between Proteus and others \citep{Banfield1992}.

\section{Conclusions and implications}\label{sec-con}
We have explored an in-situ combined formation scenario for two peculiar moons in the Neptunian system: Triton and Nereid. In our model, both moons formed in a circum-Neptunian disk, together with another set of tens of small moons. A close encounter between Neptune and an ice giant, penetrating down to the satellite system, flips the orbit of Triton and places Nereid onto a wide and eccentric orbit. Nereid's orbit immediately following this event matches observations well, but Triton's is too large and eccentric. Then, collisions between Triton and the other small moons shrink Triton's orbit on $10^4$ yr timescales. This removes the small moons and protects Nereid from elimination by Triton. Finally, tides circularise Triton's orbit over Gyrs. Our self-contained model explains the major orbital features of both Triton and Nereid.

We note that the only exomoon candidate known so far resides on an orbit tilted by $\sim(42\pm18)^\circ$ with respect to that of its host planet \citep{Teachey2018}. While further observations are needed to confirm its orbital parameters, the mechanism we propose here for solar system moons suggests an evolutionary pathway for such high-inclination satellites. Given the ubiquitousness of dynamical instability among exoplanets \citep{Rasio1996,Gong2013}, we predict a plethora of exomoons on highly-excited orbits produced during planetary encounters. Similarly, planets are themselves subject to the disturbance of stellar encounters so, as shown here for Triton and Nereid, a planet can be injected on a highly-eccentric, inclined and/or distant orbit by such events \citep{Malmberg2011,Li2019}.

\acknowledgments
{\noindent \it Acknowledgment}\\
The authors are grateful to an anonymous referee for useful comments. The authors thank Dr Craig B. Agnor for discussions and comments as well as direct text editing of the manuscript; his remarks have led to changes to the paper structure and content, including a more comprehensive discussion on collisional evolution.

D.L. thanks Anders Johansen, Douglas Hamilton, David Nesvorn\'{y} and Matija \'{C}uk for useful discussions. D.L. acknowledges the Knut and Alice Wallenberg Foundation through two grants (2014.0017, PI: Melvyn B. Davies and 2012.0150, PI: Anders Johansen). Computations were carried out at the center for scientific and technical computing at Lund University (LUNARC) through the Swedish National Infrastructure for Computing (SNIC) via project 2018/3-314. Astronomical research at the Armagh Observatory and Planetarium is funded by the Northern Ireland Department for Communities (DfC).


\end{document}